\documentclass{article}
\usepackage{spconf,amsmath,graphicx,multirow}
\usepackage{bbding}
\usepackage{booktabs}
\usepackage{cite}
\usepackage{verbatim}

\usepackage[utf8]{inputenc}
\usepackage[titletoc,title]{appendix}
\usepackage{makecell}

\newcommand{\xmark}{\ding{55}}
\newcommand{\cmark}{\ding{51}}

\usepackage{color}
\usepackage{multirow}
\usepackage{hhline}
\usepackage{pifont}
\usepackage{bm}
\usepackage{amsmath,amsfonts,amssymb,mathtools}

\usepackage{graphicx,float}

\title{EXPLOITING CROSS DOMAIN ACOUSTIC-TO-ARTICULATORY INVERTED FEATURES FOR DISORDERED SPEECH RECOGNITION}
\name{SJ Hu$^{*}$, SS Liu$^{*}$, XR Xie, MZ Geng, TZ Wang, SK Hu, MY Cui, Xunying Liu, Helen Meng
\thanks{$^*$Equal contribution. Xurong Xie is currently with Institute of Software, Chinese Academy of Sciences.}}
\address{Department of Systems Engineering and Engineering Management\\The Chinese University of Hong Kong, Hong Kong, China}

\begin{document}

\topmargin=0mm
\ninept
\setlength{\parskip}{1pt}

\maketitle

\begin{abstract}
\vspace{-0.3cm}
Articulatory features are inherently invariant to acoustic signal distortion and have been successfully incorporated into automatic speech recognition (ASR) systems for normal speech. Their practical application to disordered speech recognition is often limited by the difficulty in collecting such specialist data from impaired speakers. This paper presents a cross-domain acoustic-to-articulatory (A2A) inversion approach that utilizes the parallel acoustic-articulatory data of the 15-hour TORGO corpus in model training before being cross-domain adapted to the 102.7-hour UASpeech corpus and to produce articulatory features. Mixture density networks based neural A2A inversion models were used. A cross-domain feature adaptation network was also used to reduce the acoustic mismatch between the TORGO and UASpeech data. On both tasks, incorporating the A2A generated articulatory features consistently outperformed the baseline hybrid DNN/TDNN, CTC and Conformer based end-to-end systems constructed using acoustic features only. The best multi-modal system incorporating video modality and the cross-domain articulatory features as well as data augmentation and learning hidden unit contributions (LHUC) speaker adaptation produced the lowest published word error rate (WER) of 24.82\% on the 16 dysarthric speakers of the benchmark UASpeech task.
\end{abstract}
\begin{keywords}
Articulatory Inversion, Dysarthric Speech, Speech Recognition, Domain Adaptation
\end{keywords}
\vspace{-0.3cm}
\section{Introduction}
\vspace{-0.3cm}

Speech disorders such as dysarthria are often associated with neuro-motor conditions \cite{maas2008principles}, including cerebral palsy \cite{whitehill2000speech}, amyotrophic lateral sclerosis \cite{makkonen2018speech}, Parkinson disease \cite{scott1984evidence}, stroke or traumatic brain injuries \cite{jerntorp1992stroke}, leading to weakness or paralysis of muscles controlling articulation \cite{hixon1964restricted} 
and reduced intelligibility of speech for human listeners. They affect millions of people around the world. People with speech impairment often experience co-occurring physical disabilities and mobility issue. Their difficulty in using keyboard, mouse and touch screen based user interfaces makes speech controlled assistive technologies more natural alternatives \cite{young2010difficulties}, 
even though speech quality is degraded.
Despite the rapid progress of automatic speech recognition (ASR) technologies in the past few decades, recognition of disordered speech is still a very challenging task to date due to severe mismatch against normal speech, difficulty in large scale data collection for system development and high level of variability among speakers \cite{christensen2013combining, sehgal2015model, yu2018development,xiong2020source,liu2020exploiting, geng2022spectro}.
\par
Human speech production is a process that involves the coordinated movements of various articulators such as the tongue, lips, teeth and palate. Articulatory movement features are inherently invariant to extrinsic acoustic distortion, for example, due to environmental noise. They have been successfully applied to both normal speech \cite{zlokarnik1995adding,wrench2000continuous,kirchhoff2002combining,ghosh2011automatic,mitra2017hybrid,mitra2010articulatory} and pathological speech \cite{deng2009disordered, rudzicz2010articulatory, gonzalez2017direct, xiong2018deep, yilmaz2019articulatory,maharana2021acoustic} recognition tasks.
\par
The practical and wider use of articulatory features in ASR systems for both normal and disordered speech task domains is often limited by the difficulty in collecting sufficient quantities of such specialist data that are essential for current deep learning technologies. In practice, recording detailed articulatory movements and vocal tract shape normally requires the use of intrusive electromagnetic articulography (EMA) \cite{engwall2000static} technologies or magnetic resonance imaging (MRI) \cite{narayanan2014real}.
In the context of articulatory recordings from impaired speakers, the requirement of specialist facilities is further compounded with their underlying neuro-motor conditions, mobility issues and fatigue when speaking, leading to increasing difficulty in articulatory data collection.
\par
An alternative approach to obtain articulatory information is to estimate it from the more accessible acoustic speech signals using data driven artificial neural network based acoustic-to-articulatory (A2A) inversion \cite{papcun1992inferring, ghosh2010generalized, uria2012deep, xie2018investigation, maharana2021acoustic} 
techniques based on, for example, multilayer perceptron (MLP) \cite{papcun1992inferring} and mixture density network (MDN) \cite{ uria2012deep, xie2018investigation}. As the A2A inversion model training only requires a part of training materials to contain parallel acoustic-articulatory data, the resulting inversion model can be used to produce articulatory features when only audio recordings are available. A wider and more practical application of articulatory feature representation in ASR systems thus becomes possible. Prior researches on A2A inversion were conducted predominantly on normal speech task domains \cite{papcun1992inferring, ghosh2010generalized, uria2012deep, xie2018investigation}. In contrast, very limited researches were carried out on A2A inversion for disordered speech recognition \cite{xiong2018deep,maharana2021acoustic}. 
\par
In order to address the issues mentioned above, this paper presents a cross-domain A2A inversion approach that utilizes the parallel acoustic-articulatory data of 15-hour TORGO corpus \cite{rudzicz2012torgo} in model training before being cross-domain adapted to the 102.7-hour UASpeech corpus \cite{kim2008dysarthric} to produce articulatory features. Mixture density networks based deep neural network A2A inversion models were used. 
A cross-domain adaptation network was used to reduce the acoustic mismatch between the TORGO and UASpeech data. 
On both tasks, incorporating the generated articulatory features consistently outperformed the baseline hybrid DNN/TDNN \cite{peddinti2015time}, CTC \cite{graves2006connectionist} or Conformer \cite{gulati2020conformer
} based end-to-end systems constructed using acoustic features only. 
\par
The main contributions are summarized below. To the best of our knowledge, this work is the first use of real acoustic-articulatory parallel recordings trained A2A inversion models for articulatory features generation targeting disordered speech recognition. In contrast, related previous works either used: a) synthesized normal speech acoustic-articulatory features trained A2A inversion models before being applied to dysarthric speech \cite{xiong2018deep}, while the large mismatch between normal and impaired speech encountered during inversion model training and articulatory feature generation stages was not taken into account; or b) only considered the cross-domain or cross-corpus A2A inversion \cite{maharana2021acoustic} while the quality of generated articulatory features was not assessed using the back-end disordered speech recognition systems. In addition, the lowest published WER of 24.82\% on the benchmark UASpeech task in comparison against recent researches \cite{christensen2013combining, sehgal2015model, yu2018development, xiong2020source, liu2020exploiting, geng2022spectro, geng2020investigation, jin2021adversarial, liu2021recent} 
was obtained using the proposed cross-domain acoustic-to-articulatory inversion approach. 
\par
The rest of the paper is organized as follows. The baseline ASR systems and their incorporation of articulatory features are presented in Section 2. Section 3 presents the cross-domain A2A inversion systems. Experimental results are shown in Section 4. The conclusions are drawn and future works are discussed in Section 5. 

\vspace{-0.3cm}

\section{Articulatory Feature Based Disordered Speech Recognition}
\vspace{-0.2cm}

This section describes the time delay neural network (TDNN) \cite{peddinti2015time} 
based ASR and acoustic-articulatory feature based speech recognition (AASR) system architecture on the TORGO dataset \cite{rudzicz2012torgo} which provides parallel acoustic-articulatory data. 

\vspace{-0.4cm}
\begin{figure}[htb]
    \begin{center}
        \centerline{\includegraphics[width=6.5cm]{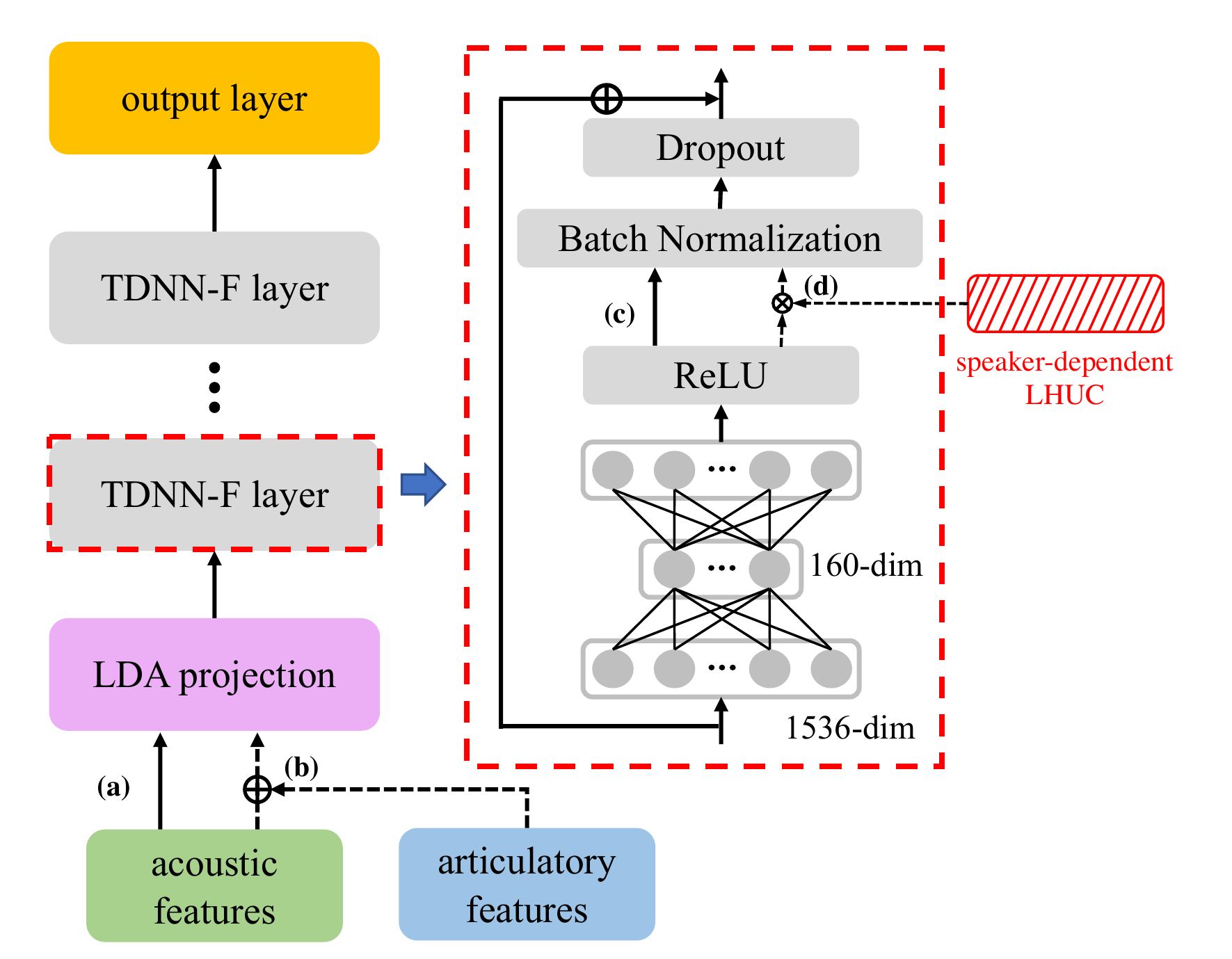}}
    \end{center}
    
    \vspace{-1cm}
\caption{The factored TDNN based ASR and AASR system architecture for the TORGO task. The AASR system uses the acoustic-articulatory concatenation via connection (b). The connection (d) is used for LHUC-SAT training for both ASR and AASR systems.}
\vspace{-0.2cm}
\end{figure}

\par
Both of the baseline ASR and AASR systems share the same main structure based on a 7-layer factorized TDNN (TDNN-F) model with a semi-orthogonal constraint 
and they were trained using the sequence discriminative lattice-free MMI (LF-MMI) criterion 
(see Figure 1). Linear discriminant analysis (LDA) based affine projection was also applied to the input acoustic only for the ASR system, or concatenated acoustic-articulatory features for the AASR system. The following TDNN-F hidden layers positioned after LDA projection are shown in the red dotted box in Figure 1. Each TDNN-F layer contains a set of neural operations performed in sequence immediately after the factorized hidden layers including: rectified linear unit (ReLU) activation, batch normalization and dropout modules. Each hidden layer’s inputs prior to context splicing were scaled and added to its outputs by a skip connection.

\par
To model the large variability among disordered speakers, learning hidden unit contributions (LHUC) \cite{swietojanski2016learning} based speaker adaptive training (SAT) was used (right part of Figure 1). Speaker-level LHUC scaling factors (in red) are applied to the ReLU activation outputs via connection (d). Supervised estimation of LHUC factors is performed for each speaker during the training stage. During test adaptation, unsupervised LHUC adaptation is used to re-estimate the LHUC scaling factors based on the speaker specific data.

\vspace{-0.3cm}
\section{Acoustic-to-articulatory inversion}
\vspace{-0.3cm}
\subsection{In-domain A2A Inversion}
\vspace{-0.2cm}
Data augmentation techniques play a vital role to address the data sparsity problem in current disordered speech recognition systems 
\cite{geng2020investigation,jin2021adversarial}. Spectral-temporal perturbation of the limited audio data collected from impaired speakers is normally used to inject more diversity into the augmented data to improve the resulting ASR system generalization on the same task, for example, the TORGO corpus. The construction of AASR systems using such augmented acoustic data requires an in-domain acoustic-to-articulatory (A2A) inversion process to produce the desired articulatory features for the expanded audio data. One of the commonly adopted neural network based A2A inversion methods is based on mixture density networks (MDNs) \cite{uria2012deep, xie2018investigation}. This is also considered in this paper. Instead of directly generating articulatory features, MDNs model the Gaussian mixture model density distribution parameters that characterise the articulatory movements. The MDN loss function is defined as 
\begin{equation}
    \setlength{\abovedisplayskip}{2pt}
    \setlength{\belowdisplayskip}{2pt}
    \mathcal L_{MDN} = - \sum_t \ln \sum_m^M {\cal S}_m({\bf y}_{t}^{\bm {\lambda}}){\cal N}({\bf a}_{t};{\bm \mu}_{t,m}, {\bm \sigma}_{t,m}^{2})
\end{equation}
where $M$ is the number of mixture components, ${\bf a}_{t}$ denotes articulatory feature vector at the $t$-th frame, $\cal S$ and $\cal N$ denote the Softmax activation and Gaussian distribution respectively, ${\bf y}_{t}^{\bm {\lambda}}$ represents the MDN network output fed into the Softmax activation to produce the mixture component weights ${\cal S}_m({\bf y}_{t}^{\bm {\lambda}})$ at time $t$. The $t$-th frame mixture component mean and variance parameters are predicted using the respective MDN outputs as ${\bm \mu}_{t,m} = {\bf y}_{t,m}^{\bm \mu}$, and ${\bm \sigma}_{t,m}^{2}= {\bf \exp}^{2}\left( {\bf y}_{t,m}^{\bm \sigma} \right)$.
The articulatory movements directly produced by MDNs usually contain artefacts. These can be further smoothed using the maximum likelihood parameter generation (MLPG) algorithm \cite{tokuda2000speech} performed on the articulatory trajectories augmented with their differentials $\Delta$ and $\Delta\Delta. $





%

\par
In this paper, a multi-task learning (MTL) approach was also adopted to construct the A2A inversion system illustrated in the right part of Figure 2, where the acoustic features were fed into the inversion model using the connection (a). Two groups of tasks including: a) an interpolation between the MDN error loss of Eqn. (1), MSE and Pearson correlation computed against the ground truth articulatory features; and b) an auxiliary monophone classification task based on phonetic alignments obtained from the HTK toolkit \cite{young2002htk} are combined in the following MTL cost function.
\begin{equation}
    \setlength{\abovedisplayskip}{2pt}
    \setlength{\belowdisplayskip}{2pt}
    \begin{matrix} \mathcal L = \underbrace{ \omega_1 \mathcal L_{MDN} + \omega_2 \mathcal L_{MSE} + \omega_3 \mathcal L_{Pearson} }_{\rm \bf regression}  + \underbrace{ \omega_4 \mathcal L_{CE} }_{\rm \bf classification} \end{matrix}
\end{equation}
where $\mathcal L_{MDN}$ is calculated by Eqn. (1), $\mathcal L_{MSE}$ and $\mathcal L_{Pearson}$ are the MSE and negative Pearson correlation coefficient calculated using the inverted and target articulatory features respectively. $\mathcal L_{CE}$ is the monophone level cross entropy (CE) loss. In this paper, the task weight parameters $\omega_1,\omega_2,\omega_3,\omega_4$ were set as the same weight 0.25\footnote{Compared with using equal weights, alternative weighting by removing the CE cost, increasing/decreasing the CE cost weight or removing one of the other three costs all led to WER increase in the resulting AASR systems.}.

\vspace{-0.3cm}
\subsection{Cross-domain A2A Inversion }
\vspace{-0.2cm}
Due to the acoustic domain mismatch, a direct cross-domain application of the A2A inversion model (described in Section 3.1) trained on the TORGO acoustic-articulatory parallel data to the UASpeech acoustic data is problematic, as was shown in the previous research on cross-domain audio-visual inversion \cite{liu2020exploiting}. To this end, the large acoustic domain mismatch between the two data sets can be minimized using multi-level adaptive networks (MLAN) \cite{bell2012transcription, liu2020exploiting,liu2021recent} before A2A inversion can be performed. An example MLAN model is shown in the left part of Figure 2 (circled in red dotted line).

\vspace{-0.3cm} 
\begin{figure}[htb]
    
\begin{minipage}[b]{1.0\linewidth}

  \centering
  \centerline{\includegraphics[width=8.5cm]{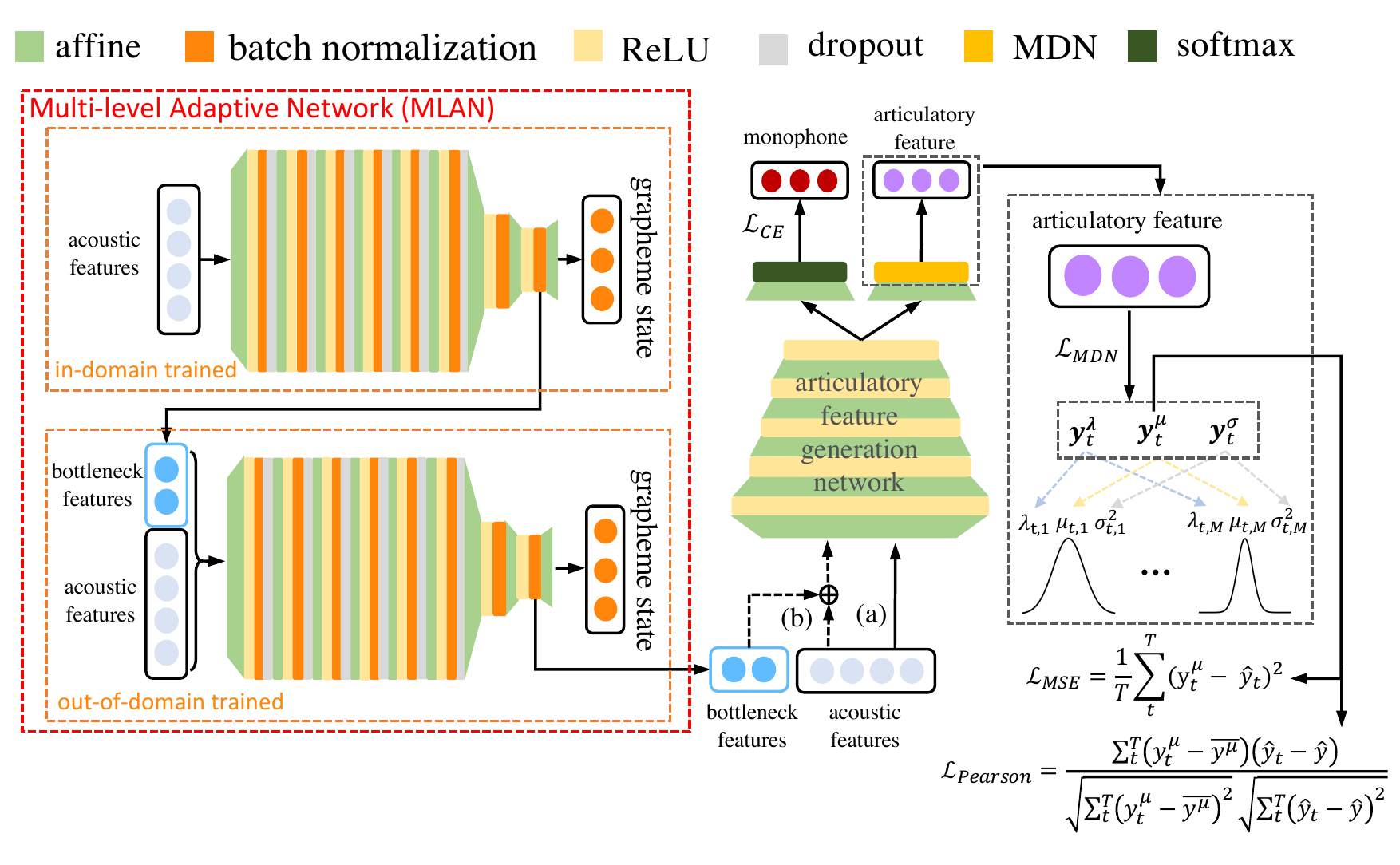}}
  \vspace{-0.4cm}

\end{minipage}

\caption{Cross-domain articulatory feature generation system architecture. The left part shows the MLAN cross domain feature adaptation network, while on the right is the multi-task trained MDN based A2A inversion model. The inputs fed into the A2A inversion model component are based on either: a) splicing windowed frames of original acoustic features using connection (a) for in-domain A2A inversion performed on TORGO acoustic data w/o data augmentation applied; or b) the MLAN adapted UASpeech bottleneck features for cross-domain A2A inversion of the UASpeech acoustic data.}
\vspace{-0.3cm}

\end{figure}

An example MLAN network consisting of two DNN components is shown in the left portion of Figure 2. Each component DNN contains a bottleneck layer positioned immediately before the output layer. The MLAN training process includes the following steps: 1) the first-level DNN was trained with the audio data from the in-domain UASpeech corpus; 2) the resulting in-domain dysarthric speech trained DNN was then used to produce bottleneck features for the out-of-domain data of the TORGO audio; 3) the second-level DNN was trained using the out-of-domain TORGO audio data concatenated with the bottleneck features computed from the previous step. When feedforwarding the UASpeech data into the resulting MLAN network, the combined effect produced by these two cascaded component DNNs is such that the final bottleneck features produced at the second-level DNN component will exhibit smaller mismatch against the bottleneck features obtained by feedforwarding the TORGO data  into the MLAN network. These cross-domain adapted bottleneck features 
are used in A2A inversion model training and articulatory feature generation (in the right part of Figure 2 using connection (b)) for the UASpeech audio data.

\vspace{-0.3cm}

\section{Experiments}
\vspace{-0.3cm}
\subsection{Experiments on the TORGO dataset}
\vspace{-0.2cm}
\textbf{Task Description and Experimental Setup :} The TORGO dataset is a disordered speech corpus with acoustic-articulatory parallel recordings and contains 8 dysarthric and 7 control speakers. In this paper, 13.49-hour audio data was used, in which the total duration of short sentence based utterances is 5.81 hours while that of single word based utterances is 7.68 hours. The total number of speech utterances is 16394. A speaker level data partitioning similar to that used in the UASpeech \cite{kim2008dysarthric} corpus is used. All 7 control speakers' data together with two thirds of the 8 dysarthric speakers' data were merged into the training set (11.7 hours) while the remaining dysarthric speech served as the test data (1.79 hours). The percentage of spoken content overlap between the training and test data is 50\%. Excessive silence at the sentence start and end was removed using a HTK toolkit \cite{young2002htk} trained GMM-HMM system. After silence stripping, the training set contains 6.46 hours of data while the test set has 1.02 hours of speech. After a combination of disordered speaker independent and dependent speed perturbation \cite{geng2020investigation} based data augmentation, the total amount of training data was expanded by a factor of 6 times and increased to 34.11 hours in total.

\par
In our experiments, the 7-layer LF-MMI based TDNN-F model as shown in Figure 1 was implemented using the Kaldi toolkit \cite{povey2011kaldi} while Conformer based end-to-end systems\footnote{8 encoder plus 4 decoder layers, feed-forward layer dim = 1024, attention heads = 4, dim of attention heads = 256, interpolated CTC+AED cost.} were implemented using the Espnet toolkit \cite{watanabe2018espnet}. A 3-frame context window was used in both ASR and AASR hybrid LF-MMI trained TDNN systems. 40-dimension Mel-scale filter banks (FBKs) were used as the input acoustic features, and the articulatory features were represented by the measured articulatory trajectories. 6 trajectory variables (TV) were selected as articulatory features, i.e. tongue tip (TT), tongue middle (TM), tongue back (TB), upper lip (UL), lower lip (LL) and lower incisor (LI). The X, Y and Z coordinates which capture the spatial movement of the measured articulators were used to construct 18-dimension articulatory feature vectors before being concatenated with FBK features. A 3-gram LM trained by all the TORGO transcripts with a vocabulary size of 1578 words was used in decoding.

\vspace{-0.5cm}
\begin{table}[H]
\caption{Comparison of the WER results produced by various ASR and AASR systems on the 8 TORGO dysarthric speakers test set. The dysarthric speakers are grouped by their intelligibility levels, i.e. ``Severe'', ``Moderate'' and ``Mild''. ``Aug.'' and ``arti.'' are the abbreviations of augmentation and articulatory respectively. AA fusion includes different AA modality fusion: A) input feature concatenation; AND B) score fusion. $^\dag$ and $^\ddag$ denote a statistically significant improvement compared with baseline and augmented ASR systems respectively (Sys. 1 and Sys. 6).}
\vspace{0.1cm}
\centering
\scalebox{0.63}{
\begin{tabular}{c|c|c|c|c|c|c|c|c|c} 
\hline\hline
\multirow{2}{*}{Sys.} & \multirow{2}{*}{Model}                                               & \multirow{2}{*}{\begin{tabular}[c]{@{}c@{}}arti.\\source\end{tabular}} & \multirow{2}{*}{\begin{tabular}[c]{@{}c@{}}Data\\Aug.\end{tabular}} & \multirow{2}{*}{\begin{tabular}[c]{@{}c@{}}LHUC\\SAT\end{tabular}}   & \multirow{2}{*}{AA Fusion} & \multicolumn{4}{c}{WER \%}                        \\ 
\cline{7-10}
                      &                                                                      &                                                                        &                                                                     &                          &                            & Severe & Moderate & Mild & Average                \\ 
\hline\hline
1                     & \multirow{5}{*}{TDNN} & \xmark                                                                      & \xmark                                                                   & \xmark                        & \xmark                          & 16.22  & 10.31    & 3.87 & 11.62                  \\ 
\cline{1-1}\cline{3-10}
2                     &                                                                      & \multirow{2}{*}{original}                                              & \multirow{2}{*}{\xmark}                                                  & \multirow{2}{*}{\xmark}       & input                      & 15.00  & 9.59     & 4.18 & 10.93                  \\
3                     &                                                                      &                                                                        &                                                                     &                          & score A+AA                 & 13.98  & 9.39     & 3.79 & \textbf{10.25}$^\dag$  \\ 
\cline{1-1}\cline{3-10}
4                     &                                                                      & \multirow{2}{*}{inversion}                                             & \multirow{2}{*}{\xmark}                                                  & \multirow{2}{*}{\xmark}       & input                      & 15.61  & 10.10    & 3.79 & 11.24                  \\
5                     &                                                                      &                                                                        &                                                                     &                          & score A+AA                 & 14.67  & 9.80     & 3.41 & \textbf{10.59}$^\dag$  \\ 
\hline\hline
6                     & \multirow{5}{*}{TDNN} & \xmark                                                                      & \cmark                                                             & \xmark                        & \xmark                          & 12.80  & 8.78     & 3.64 & 9.47                   \\ 
\cline{1-1}\cline{3-10}
7                     &                                                                      & \multirow{2}{*}{original}                                              & \multirow{2}{*}{\cmark}                                            & \multirow{2}{*}{\xmark}       & input                      & 13.25  & 7.24     & 3.02 & 9.21                   \\
8                     &                                                                      &                                                                        &                                                                     &                          & score A+AA                 & 12.68  & 7.76     & 2.86 & 8.98                   \\ 
\cline{1-1}\cline{3-10}
9                     &                                                                      & \multirow{2}{*}{inversion}                                             & \multirow{2}{*}{\cmark}                                            & \multirow{2}{*}{\xmark}       & input                      & 12.72  & 7.45     & 3.41 & 9.09                   \\
10                    &                                                                      &                                                                        &                                                                     &                          & score A+AA                 & 12.28  & 7.96     & 2.86 & \textbf{8.81}$^\ddag$  \\ 
\hline\hline
11                    & \multirow{3}{*}{TDNN} & \xmark                                                                      & \cmark                                                             & \cmark                  & \xmark                          & 12.52  & 8.27     & 3.25 & 9.11                   \\ 
\cline{1-1}\cline{3-10}
12                    &                                                                      & \multirow{2}{*}{inversion}                                             & \multirow{2}{*}{\cmark}                                            & \multirow{2}{*}{\cmark} & input                      & 12.52  & 7.14     & 3.17 & 8.85                   \\
13                    &                                                                      &                                                                        &                                                                     &                          & score A+AA                 & 12.03  & 7.35     & 2.94 & 8.58                   \\ 
\hline\hline
14                    & \multirow{2}{*}{\begin{tabular}[c]{@{}c@{}}Con-\\former\end{tabular}}                                                 & \xmark                                                                      & \cmark                                          & \xmark                        & \xmark                          & 22.28   & 7.35     & 4.72  & 14.39                   \\ 
\cline{1-1}\cline{3-10}
15                    &                                                                      & inversion                                                              & \cmark                                          & \xmark                        & input                      & 21.79   & 7.55      & 3.79  & 13.93                   \\
\hline\hline
\end{tabular}
}
\vspace{-0.3cm}
\end{table}

\noindent
\textbf{Results :} The performance\footnote{A matched pairs sentence-segment word error based statistical significance test was performed at a significance level $\alpha$ = 0.05} of various ASR and acoustic-articulatory feature based AASR systems on the 8 TORGO dysarthric speakers test set is shown in Table 1. Several trends can be found: \textbf{1)} the incorporation of the original articulatory features provided in the TORGO data consistently outperform the acoustic only TDNN ASR systems with or without data augmentation (Sys. 2 \& 3 vs. Sys. 1; Sys. 7 \& 8 vs. Sys. 6); among these, statistically significant WERs reduction of 1.37\% absolute (11.79\% relative) and 1.03\% absolute (8.86\% relative) were obtained over the baseline ASR system using the AASR system Sys. 3 (original articulatory features) and Sys. 5 (inverted articulatory features) respectively. The RMSE of the MDN based A2A inversion system for inverted features in Sys. 5 is 0.808mm. \textbf{2)} further score fusion between the ASR and AASR systems consistently outperforms the AASR systems using the acoustic-articulatory feature concatenation at the input (Sys. 3 vs Sys. 2; Sys. 8 vs. Sys. 7). \textbf{3)} The AASR systems trained using A2A inverted features produced performance comparable to those using the original articulatory features (Sys. 9 \& 10 vs. Sys. 7 \& 8) after data augmentation. \textbf{4)} Consistent WER reductions over the baseline ASR systems were obtained using the AASR systems constructed using the A2A inverted articulatory features before and after LHUC-SAT speaker adaptation (Sys. 10 vs. Sys. 6; Sys. 13 vs. Sys. 11). \textbf{5)} Similar trends can also be found on the Conformer based systems (Sys. 15 vs. Sys. 14). No  score fusion was used on Conformer due to no performance gain compared to TDNN systems.

\vspace{-0.3cm}
\subsection{Experiments on the UASpeech dataset}
\vspace{-0.2cm}

\noindent
\textbf{Task Description and Experimental Setup :} The UASpeech corpus is the largest publicly available disordered speech corpus that is designed as an isolated word recognition task \cite{kim2008dysarthric} consisting of 16 dysarthric and 13 control speakers. The speech materials contain 155 common words and 300 uncommon words. The entire corpus is further divided into 3 subset blocks per speaker, with each block containing all 155 common words and one third of the uncommon words. The data from Block 1 (B1) and Block 3 (B3) of all the 29 speakers are used as the training set (69.1 hours of audio, 99195 utterances in total), while the data of Block 2 (B2) collected from all the 16 dysarthric speakers serves as the evaluation data set (22.6 hours of audio, 26520 utterances in total). After removing excessive silence at the start and end of speech audio segments, a combined total of 30.6 hours of audio data from Block 1 and 3 were used as the training set, while 9 hours of speech from Block 2 was used for performance evaluation. After speaker independent and dependent speed perturbation based data augmentation \cite{geng2020investigation, liu2021recent}, the total amount of training data was increased to 130.1 hours in total.
\par

\vspace{-0.5cm}
\begin{table}[H]
\centering
\caption{WERs of baseline ASR and AASR systems using the cross-domain inverted articulatory features on the UASpeech test set of 16 dysarthric speakers grouped by intelligibility levels: ``Very low'', ``Low'', ``Mild'' and ``High''. AA fusion includes different AA modality fusion: A) input feature concatenation; B) hidden layer fusion; AND C) score fusion. Optional further incorporation of video modality is used. $\dag$ denotes statistical significant differences obtained against the baseline ASR systems (Sys. 1, 7, 12 \& 14)}
\scalebox{0.57}{
\begin{tabular}{c|c|c|c|c|c|c|c|c|c|c} 
\hline\hline
\multirow{2}{*}{Sys.} & \multirow{2}{*}{model}                                                & \multirow{2}{*}{AA Fusion}  & \multirow{2}{*}{\begin{tabular}[c]{@{}c@{}}Data\\Aug.\end{tabular}} & \multirow{2}{*}{\begin{tabular}[c]{@{}c@{}}LHUC\\SAT\end{tabular}} & \multirow{2}{*}{MLAN}  & \multicolumn{5}{c}{WER \%}                                                                                            \\ 
\cline{7-11}
                      &                                                                       &                             &                                                                     &                                                                    &                        & Very Low              & Low                   & Mild                  & High                 & Average                \\ 
\hline\hline
1                     & \multirow{10}{*}{\begin{tabular}[c]{@{}c@{}}Hybird\\DNN\end{tabular}} & \xmark                       & \multirow{3}{*}{\xmark}                                              & \multirow{6}{*}{\xmark}                                             & \multirow{2}{*}{\xmark} & 69.82                 & 32.61                 & 24.53                 & 10.40                & 31.45                  \\ 
\cline{1-1}\cline{3-3}\cline{7-11}
2                     &                                                                       & \multirow{2}{*}{7th hidden} &                                                                     &                                                                    &                        & 69.41                 & 33.01                 & 24.52                 & 10.35                & 31.46                  \\ 
\cline{6-11}
3                     &                                                                       &                             &                                                                     &                                                                    & \cmark                  & \textbf{67.82}$^\dag$ & \textbf{31.25}$^\dag$ & \textbf{22.88}$^\dag$ & \textbf{9.77}$^\dag$ & \textbf{30.15}$^\dag$  \\ 
\cline{1-1}\cline{3-4}\cline{6-11}
4                     &                                                                       & \xmark                       & \multirow{3}{*}{\cmark}                                              &                                                                    & \multirow{2}{*}{\xmark} & 66.84                 & 28.70                 & 20.39                 & 9.37                 & 28.67                  \\ 
\cline{1-1}\cline{3-3}\cline{7-11}
5                     &                                                                       & \multirow{2}{*}{7th hidden} &                                                                     &                                                                    &                        & 66.43                 & 30.26                 & 22.13                 & 9.63                 & 29.41                  \\ 
\cline{6-11}
6                     &                                                                       &                             &                                                                     &                                                                    & \cmark                  & 66.27                 & 28.42                 & 19.78                 & 9.40                 & \textbf{28.37}         \\ 
\cline{1-1}\cline{3-11}
7                     &                                                                       & \xmark                       & \multirow{3}{*}{\cmark}                                              & \multirow{5}{*}{\cmark}                                             & \multirow{2}{*}{\xmark} & 64.22                 & 27.87                 & 18.01                 & 7.60                 & 26.85                  \\ 
\cline{1-1}\cline{3-3}\cline{7-11}
8                     &                                                                       & \multirow{2}{*}{7th hidden} &                                                                     &                                                                    &                        & 62.40                 & 27.81                 & 18.64                 & 8.69                 & 26.93                  \\ 
\cline{6-11}
9                     &                                                                       &                             &                                                                     &                                                                    & \cmark                  & \textbf{62.08}$^\dag$ & \textbf{26.23}$^\dag$ & 17.33                 & 8.47                 & \textbf{26.13}$^\dag$  \\ 
\cline{1-1}\cline{3-4}\cline{6-11}
10                    &                                                                       & score A+AA                  & \cmark                                                               &                                                                    & \cmark                  & \textbf{61.22}$^\dag$ & \textbf{25.51}$^\dag$ & \textbf{16.15}$^\dag$ & 7.64                 & \textbf{25.26}$^\dag$  \\ 
\cline{1-4}\cline{6-11}
11                    & +visual                                                               & score A+AA                  & \cmark                                                               &                                                                    & \cmark                  & \textbf{60.14}        & \textbf{25.11}        & \textbf{15.92}        & \textbf{7.46}        & \textbf{24.82}         \\ 
\hline\hline
12                    & \multirow{2}{*}{CTC}                                                  & \xmark                       & \multirow{2}{*}{\cmark}                                              & \multirow{4}{*}{\xmark}                                             & \multirow{4}{*}{\cmark} & 78.25                 & 53.18                 & 46.56                 & 34.09                & 50.79                  \\
13                    &                                                                       & input                       &                                                                     &                                                                    &                        & \textbf{74.33}$^\dag$ & 52.53                 & 47.13                 & 34.49                & \textbf{50.04}$^\dag$  \\ 
\cline{1-4}\cline{7-11}
14                    & \multirow{2}{*}{\begin{tabular}[c]{@{}c@{}}Con-\\former\end{tabular}} & \xmark                       & \multirow{2}{*}{\cmark}                                              &                                                                    &                        & 68.22                 & 49.38                 & 47.51                 & 41.77                & 50.44                  \\
15                    &                                                                       & input                       &                                                                     &                                                                    &                        & 67.42                 & \textbf{48.53}$^\dag$ & 47.20                 & 41.95                & \textbf{50.05}         \\
\hline\hline
\end{tabular}
}
\end{table}

\vspace{-0.3cm}
\noindent
\textbf{Results :} Domain adaptation is essential in transferring articulatory features from TORGO corpus to UASpeech dysarthric corpus (Sys. 3 vs. Sys. 2; Sys. 6 vs. Sys. 5; Sys. 9 vs. Sys. 8).
After incorporating the corss-domain inverted articulatory features obtained on the UASpeech audio data, 
 the resulting AASR systems shown in Table 2 consistently outperform the comparable baseline ASR systems using acoustic feature only by a significant margin in WER (Sys. 3 vs. Sys. 1; Sys. 6 vs. Sys. 4; Sys. 9 vs. Sys. 7) before and after data augmentation and LHUC-SAT based speaker adaptation applied to a state-of-the-art hybird DNN system \cite{liu2021recent}, a CTC end-to-end system (Sys. 13 vs. Sys. 12) and a Conformer end-to-end system (Sys. 15 vs. Sys. 14). In particular, significant WER reductions of 1.64\%-2.14\% were 
 obtained on the ``Very low'' and ``Low'' subgroups after data augmentation and LHUC-SAT (Sys. 9 vs. Sys. 7).
\par
Further ablation studies were conducted to investigate alternative forms of acoustic-articulatory modality fusion. Table 2 shows the performance of 130.1-hour augmented training data based LHUC-SAT speaker adapted baseline acoustic feature only ASR system (Sys. 7), and two acoustic-articulatory features based AASR systems constructed using either a) 7-th hidden layer feature fusion (Sys. 9), or b) score interpolation (Sys. 10, equal weighting to ASR and AASR systems) based acoustic-articulatory modality fusion, with an optional further incorporation of video modality (Sys. 11).

\par
Score fusion between the ASR and AASR systems produced a statistically significant WER reduction of 1.59\% over the comparable ASR system (Sys. 10 vs. Sys. 7) and 0.87\% over the AASR system using hidden layer feature fusion (Sys. 10 vs. Sys. 9), although with increased system complexity. The lowest WER of 24.82\% was obtained by further incorporating visual features \cite{liu2021recent} (Sys. 11). To the best of our knowledge, this is the lowest WER published so far on the UASpeech test set of 16 dysarthric speakers reported in the literature. Performance of this system against previously published systems on the same task are shown in Table 3.

\vspace{-0.6cm}
\begin{table}[H]
\caption{A comparison between published systems on UASpeech and our system. ``DA'' stands for data augmentation.}
\vspace{-0.3cm}
\begin{center}
\scalebox{0.8}{
\begin{tabular}{cc} 
\hline\hline
systems                                                  & WER\%  \\ 
\hline
Sheffield-2013 Cross domain augmentation \cite{christensen2013combining}                 & 37.50  \\
Sheffield-2015 Speaker adaptive training \cite{sehgal2015model}                 & 34.80  \\
CUHK-2018 DNN System Combination \cite{yu2018development}                         & 30.60  \\
Sheffield-2020 Fine-tuning CNN-TDNN speaker adaptation \cite{xiong2020source}   & 30.76  \\
CUHK-2020 Cross-domain AVSR \cite{liu2020exploiting}                              & 26.84  \\
CUHK-2020 DNN + DA + LHUC SAT \cite{geng2020investigation}                            & 26.37  \\
CUHK-2021 DNN + GAN based DA + LHUC SAT \cite{jin2021adversarial}                  & 25.89  \\
CUHK-2021 NAS DNN + DA + LHUC SAT + AV fusion \cite{liu2021recent}            & 25.21  \\
\textbf{DNN + DA + LHUC SAT + AAV fusion (ours)}         & \textbf{24.82}  \\
\hline\hline
\end{tabular}
}
\vspace{-0.8cm}
\end{center}
\end{table}

\vspace{-0.4cm}
\section{Conclusion}
\vspace{-0.3cm}
This paper presents a cross-domain acoustic-to-articulatory (A2A) inversion approach that utilizes small amounts of parallel acoustic-articulatory data of the 15-hour TORGO corpus in model training before being cross-domain adapted to a larger 102.7-hour UASpeech dysarthric corpus to produce articulatory features for ASR system construction incorporating articulatory features. Experimental results on both tasks suggest that incorporating the A2A generated articulatory features consistently outperformed the baseline hybrid DNN/TDNN, CTC and Conformer based end-to-end systems constructed using acoustic features only, while producing the lowest published WER of 24.82\% on the 16 dysarthric speakers of the benchmark UASpeech task. The proposed cross-domain A2A inversion method allows a more practical and wider use of articulatory features in ASR systems targeting disordered speech.

\vspace{-0.3cm}

\section{Acknowledgement}
\vspace{-0.3cm}
This research is supported by Hong Kong RGC GRF grant No. 14200218, 14200220, TRS T45-407/19N, ITF grant No. ITS/254/19, SHIAE grant No. MMT-p1-19 and National Natural Science Foundation of China (NSFC) Grant 62106255.

\vspace{-0.3cm}
\bibliographystyle{IEEEbib}
\bibliography{refs}

\begin{thebibliography}{10}

\bibitem{maas2008principles}
Edwin Maas et~al.,
\newblock ``{Principles of motor learning in treatment of motor speech
  disorders},''
\newblock {\em ASHA}, 2008.

\bibitem{whitehill2000speech}
Tara~L Whitehill et~al.,
\newblock ``{Speech errors in Cantonese speaking adults with cerebral palsy},''
\newblock {\em Clinical linguistics \& phonetics}, 2000.

\bibitem{makkonen2018speech}
Tanja Makkonen et~al.,
\newblock ``{Speech deterioration in amyotrophic lateral sclerosis (ALS) after
  manifestation of bulbar symptoms},''
\newblock {\em IJLCD}, 2018.

\bibitem{scott1984evidence}
Sophie Scott et~al.,
\newblock ``{Evidence for an apparent sensory speech disorder in Parkinson's
  disease.},''
\newblock {\em JNNP}, 1984.

\bibitem{jerntorp1992stroke}
Peter Jerntorp and Goran Berglund,
\newblock ``{Stroke registry in Malm{\"o}, Sweden.},''
\newblock {\em STROKE}, 1992.

\bibitem{hixon1964restricted}
Thomas~J Hixon et~al.,
\newblock ``{Restricted motility of the speech articulators in cerebral
  palsy},''
\newblock {\em JSHD}, 1964.

\bibitem{young2010difficulties}
Victoria Young et~al.,
\newblock ``{Difficulties in automatic speech recognition of dysarthric
  speakers and implications for speech-based applications used by the elderly:
  A literature review},''
\newblock {\em ASSIST TECHNOL}, 2010.

\bibitem{christensen2013combining}
Heidi Christensen et~al.,
\newblock ``{Combining in-domain and out-of-domain speech data for automatic
  recognition of disordered speech.},''
\newblock in {\em INTERSPEECH}, 2013.

\bibitem{sehgal2015model}
Siddharth Sehgal et~al.,
\newblock ``{Model adaptation and adaptive training for the recognition of
  dysarthric speech},''
\newblock in {\em SLPAT}, 2015.

\bibitem{yu2018development}
Jianwei Yu et~al.,
\newblock ``{Development of the CUHK Dysarthric Speech Recognition System for
  the UASpeech Corpus.},''
\newblock in {\em INTERSPEECH}, 2018.

\bibitem{xiong2020source}
Feifei Xiong et~al.,
\newblock ``{Source domain data selection for improved transfer learning
  targeting dysarthric speech recognition},''
\newblock in {\em ICASSP}. IEEE, 2020.

\bibitem{liu2020exploiting}
Shansong Liu et~al.,
\newblock ``{Exploiting Cross-Domain Visual Feature Generation for Disordered
  Speech Recognition.},''
\newblock in {\em INTERSPEECH}, 2020.

\bibitem{geng2022spectro}
Mengzhe Geng et~al.,
\newblock ``{Spectro-Temporal Deep Features for Disordered Speech Assessment
  and Recognition},''
\newblock in {\em INTERSPEECH}, 2021.

\bibitem{zlokarnik1995adding}
Igor Zlokarnik,
\newblock ``{Adding articulatory features to acoustic features for automatic
  speech recognition},''
\newblock {\em JASA}, 1995.

\bibitem{wrench2000continuous}
Alan Wrench et~al.,
\newblock ``{Continuous speech recognition using articulatory data},''
\newblock in {\em ICSLP}, 2000.

\bibitem{kirchhoff2002combining}
Katrin Kirchhoff et~al.,
\newblock ``{Combining acoustic and articulatory feature information for robust
  speech recognition},''
\newblock {\em Speech Communication}, 2002.

\bibitem{ghosh2011automatic}
Prasanta~Kumar Ghosh et~al.,
\newblock ``{Automatic speech recognition using articulatory features from
  subject-independent acoustic-to-articulatory inversion},''
\newblock {\em JASA}, 2011.

\bibitem{mitra2017hybrid}
Vikramjit Mitra et~al.,
\newblock ``{Hybrid convolutional neural networks for articulatory and acoustic
  information based speech recognition},''
\newblock {\em Speech Communication}, 2017.

\bibitem{mitra2010articulatory}
Vikramjit Mitra et~al.,
\newblock ``{Articulatory information for noise robust speech recognition},''
\newblock {\em IEEE TASLP}, 2010.

\bibitem{deng2009disordered}
Yunbin Deng et~al.,
\newblock ``{Disordered speech recognition using acoustic and sEMG signals},''
\newblock in {\em INTERSPEECH}, 2009.

\bibitem{rudzicz2010articulatory}
Frank Rudzicz,
\newblock ``{Articulatory knowledge in the recognition of dysarthric speech},''
\newblock {\em IEEE TASLP}, 2010.

\bibitem{gonzalez2017direct}
Jose~A Gonzalez et~al.,
\newblock ``{Direct speech reconstruction from articulatory sensor data by
  machine learning},''
\newblock {\em IEEE TASLP}, 2017.

\bibitem{xiong2018deep}
Feifei Xiong et~al.,
\newblock ``{Deep learning of articulatory-based representations and
  applications for improving dysarthric speech recognition},''
\newblock in {\em ITG-Symposium}, 2018.

\bibitem{yilmaz2019articulatory}
Emre Y{\i}lmaz et~al.,
\newblock ``{Articulatory and bottleneck features for speaker-independent ASR
  of dysarthric speech},''
\newblock {\em Computer Speech \& Language}, 2019.

\bibitem{maharana2021acoustic}
Sarthak~Kumar Maharana et~al.,
\newblock ``{Acoustic-to-Articulatory Inversion for Dysarthric Speech by Using
  Cross-Corpus Acoustic-Articulatory Data},''
\newblock in {\em ICASSP}. IEEE, 2021.

\bibitem{engwall2000static}
Olov Engwall,
\newblock ``{Are static MRI measurements representative of dynamic speech?
  Results from a comparative study using MRI, EPG and EMA},''
\newblock in {\em ICSLP}, 2000.

\bibitem{narayanan2014real}
Shrikanth Narayanan et~al.,
\newblock ``{Real-time magnetic resonance imaging and electromagnetic
  articulography database for speech production research (TC)},''
\newblock {\em JASA}, 2014.

\bibitem{papcun1992inferring}
George Papcun et~al.,
\newblock ``{Inferring articulation and recognizing gestures from acoustics
  with a neural network trained on x-ray microbeam data},''
\newblock {\em JASA}, 1992.

\bibitem{ghosh2010generalized}
Prasanta~Kumar Ghosh et~al.,
\newblock ``{A generalized smoothness criterion for acoustic-to-articulatory
  inversion},''
\newblock {\em JASA}, 2010.

\bibitem{uria2012deep}
Benigno Uria et~al.,
\newblock ``{Deep architectures for articulatory inversion},''
\newblock in {\em INTERSPEECH}, 2012.

\bibitem{xie2018investigation}
Xurong Xie et~al.,
\newblock ``{Investigation of stacked deep neural networks and mixture density
  networks for acoustic-to-articulatory inversion},''
\newblock in {\em ISCSLP}. IEEE, 2018.

\bibitem{rudzicz2012torgo}
Frank Rudzicz et~al.,
\newblock ``{The TORGO database of acoustic and articulatory speech from
  speakers with dysarthria},''
\newblock {\em Language Resources and Evaluation}, 2012.

\bibitem{kim2008dysarthric}
Heejin Kim et~al.,
\newblock ``{Dysarthric speech database for universal access research},''
\newblock in {\em INTERSPEECH}, 2008.

\bibitem{peddinti2015time}
Vijayaditya Peddinti et~al.,
\newblock ``{A time delay neural network architecture for efficient modeling of
  long temporal contexts},''
\newblock in {\em INTERSPEECH}, 2015.

\bibitem{graves2006connectionist}
Alex Graves et~al.,
\newblock ``{Connectionist temporal classification: labelling unsegmented
  sequence data with recurrent neural networks},''
\newblock in {\em ICML}, 2006.

\bibitem{gulati2020conformer}
Anmol Gulati et~al.,
\newblock ``{Conformer: Convolution-augmented Transformer for Speech
  Recognition},''
\newblock in {\em INTERSPEECH}, 2020.

\bibitem{geng2020investigation}
Mengzhe Geng et~al.,
\newblock ``{Investigation of Data Augmentation Techniques for Disordered
  Speech Recognition.},''
\newblock in {\em INTERSPEECH}, 2020.

\bibitem{jin2021adversarial}
Zengrui Jin et~al.,
\newblock ``{Adversarial Data Augmentation for Disordered Speech
  Recognition},''
\newblock in {\em INTERSPEECH}, 2021.

\bibitem{liu2021recent}
Shansong Liu et~al.,
\newblock ``{Recent Progress in the CUHK Dysarthric Speech Recognition
  System},''
\newblock {\em IEEE TASLP}, 2021.

\bibitem{swietojanski2016learning}
Pawel Swietojanski et~al.,
\newblock ``{Learning hidden unit contributions for unsupervised acoustic model
  adaptation},''
\newblock {\em IEEE TASLP}, 2016.

\bibitem{tokuda2000speech}
Keiichi Tokuda et~al.,
\newblock ``{Speech parameter generation algorithms for HMM-based speech
  synthesis},''
\newblock in {\em ICASSP}. IEEE, 2000.

\bibitem{young2002htk}
Steve Young et~al.,
\newblock ``{The HTK book},''
\newblock {\em Cambridge university engineering department}, 2002.

\bibitem{bell2012transcription}
Peter Bell et~al.,
\newblock ``{Transcription of multi-genre media archives using out-of-domain
  data},''
\newblock in {\em SLT}. IEEE, 2012, pp. 324--329.

\bibitem{povey2011kaldi}
Daniel Povey et~al.,
\newblock ``{The Kaldi speech recognition toolkit},''
\newblock in {\em ASRU}. IEEE, 2011.

\bibitem{watanabe2018espnet}
Shinji Watanabe et~al.,
\newblock ``{Espnet: End-to-end speech processing toolkit},''
\newblock {\em arXiv:1804.00015}, 2018.

\end{thebibliography}

\end{document}